# Charge Transport Framework for Crystalline π-Conjugated Materials

*Nisarg Trivedi, Maximilian F.X. Dorfner, Michel Panhans, Frank Ortmann**

N. Trivedi, M. F.X. Dorfner, M. Panhans, F. Ortmann

Department of Chemistry, TUM School of Natural Sciences, and Atomistic Modeling Center, Munich Data Science Institute, Technical University of Munich, Germany

E-mail: frank.ortmann@tum.de

**Abstract:**

**Quantitative charge-transport modelling in crystalline π-conjugated polymers requires an electronic representation that captures extended conjugation and material-specific electron–phonon interactions on equal footing. We present a Wannier-orbital–based formulation of the transient-polaron-localization (TPL) framework for quantitative charge-transport simulations in crystalline π-conjugated polymers. By expressing all**

**interactions directly in a Wannier representation derived from first-principles calculations, the approach avoids artificial fragmentation of polymer backbones and enables a consistent, material-specific description of electronic couplings and electron–phonon interactions. Low- and high-frequency vibrational modes are incorporated within a mode-resolved TPL formalism, yielding an effective electronic Hamiltonian that captures both dynamic disorder and polaronic renormalization. The methodology is demonstrated for the ambipolar crystalline naphthalenediimide-bithiophene co-polymer, enabling a direct comparison of electron and hole transport along the polymer backbone and $\pi$-stacking directions. The simulations reproduce key transport features, including pronounced anisotropy and higher electron than hole mobility. Additional studies on single- and bi-chain systems reveal the essential role of inter-chain coherence in supporting efficient transport along the polymer backbone. Beyond this specific system, the Wannier-orbital-based TPL framework provides a transferable route to disentangle intrinsic charge-transport mechanisms in crystalline $\pi$-conjugated materials and establishes a foundation for future work.**

## 1. Introduction

Organic semiconductors (OSCs)[1] have emerged as versatile low-cost materials for a wide range of electronic and optoelectronic applications, including organic light-emitting diodes[2–5] (OLEDs), field-effect transistors[6–9] (OFETs), organic photovoltaics[10–12] (OPVs), and organic thermoelectrics.[13–15] They combine desirable features such as low-temperature fabrication, environmentally friendly nature, and mechanical flexibility. Despite their commercial success in OLED applications, OSCs typically exhibit lower carrier mobility than their inorganic counterparts, which limits their broader commercial adoption.[8] Improving charge transport, therefore, requires a detailed understanding of how molecular structure controls electronic properties. Even minor chemical modifications can reshape morphology and crystal packing[16] and, in turn, electronic properties[16–18] and charge-carrier mobility.[19–22]

Charge transport in OSCs is particularly difficult to describe theoretically[1] because several energies are of similar magnitude: transfer integrals between neighboring molecules, thermal energy scale, disorder as well as electron-phonon coupling energies.[23] Several transport models have been developed for OSCs, including incoherent hopping models for disordered molecular systems,[24,25] while early polaronic models[26,27] and surface-hopping models[28] can capture coherent transport features of organic crystals. The transient localization (TL) model[29] describes the coherent short-time localization of charge carriers induced by thermal lattice vibrations and has successfully reproduced key experimental signatures of charge dynamics in high-mobility OSCs such as rubrene.[30–32] More recently, exact quantum dynamics simulations have sought to clarify the conditions under which the TL scenario can be reliable[33,34] and when polaronic effects occur.[33] Hutsch et al.[35] used a hybrid approach combining the elements of TL theory and polaron theory, hence termed transient polaron localization (TPL) herein, achieving accurate predictions of carrier mobilities across a large and diverse set of organic crystals. Similar methods have been applied to polymer chains.[36] Non-adiabatic molecular dynamics simulations based on a first-principles–parameterized Holstein–Peierls Hamiltonian have illustrated how torsional fluctuations and energetic offsets can influence intrachain charge transport.[37] Complementarily, DFT-based ab initio molecular dynamics revealed band-like intrachain transport in PEDOT chains.[38] Transient photocurrent studies on donor–acceptor copolymers have revealed pronounced mobility relaxation governed by multiple trapping in exponential trap distributions,[39] highlighting the relevance of multiscale descriptions that connect electronic structure, morphology, and macroscopic transport in disordered bulk polymer systems.[36] In contrast, ordered crystalline polymers with well-defined packing motifs provide access to the intrinsic interplay of electronic couplings and vibrational effects, without the dominant influence of structural disorder. Despite all previous methodological progress, a consistent microscopic framework for crystalline π-conjugated polymers remains lacking.

A central methodological issue in atomistic charge-transport modeling is the choice of electronic basis, because it determines how faithfully the relevant electronic states and couplings can be represented in a tractable finite description. While Bloch states are the natural eigenstates of periodic crystals, their delocalized character makes them less suitable as a basis for transport models that typically resolve local couplings and vibrational fluctuations in real-space. Fragment-orbital approaches[40] provide such a localized representation and have been successfully applied.[28,36] While this approach is well-suited to molecular crystals, it can become problematic for extended conjugated polymers. Here, fragment-based basis functions necessarily involve artificial bond breaking and chemical termination. This disrupts the covalent bonding and π-conjugation that are intrinsic to the polymer electronic structure, such that orbitals of isolated fragments no longer ensure an adequate representation of repeat units in the polymer backbone.

In this work, we develop an accurate and efficient formulation of the TPL approach[35] using a Wannier-orbital representation[41] and extend its application to crystalline $\pi$-conjugated polymers. We demonstrate the functionality of this Wannier-orbital-based TPL approach by applying it to the benchmark polymer P(NDI2OD-T2).[42] We study the transport characteristics of different charge carriers, including their anisotropies in transport, and the relative contributions of inter- and intra-chain transport in this material. The results reproduce key experimental trends and, at the same time, establish a consistent and broadly applicable framework for charge transport in crystalline π-conjugated materials.

## 2. Results

### 2.1. Theory and Implementation

#### *2.1.1. Theoretical Framework*

In this section, we briefly summarize the TPL framework adopted here. The central idea of TPL is the separation of vibrational modes according to their characteristic time scales: low-frequency modes that effectively act as quasi-static disorder on the time scale of the charge motion, whereas high-frequency modes, faster than the transport-time scale, are treated in the polaronic limit. The separation in slow and fast modes is established by a the reference energy that is defined by the larger of the maximum electronic transfer integral $\varepsilon_{MN}^{max}$ and the thermal energy $2k_BT$.[35] This separation yields an effective electronic Hamiltonian in which vibrational effects enter through renormalized electronic parameters.

The starting point is the Holstein-Peierls Hamiltonian[43,44], which includes terms for electronic, phononic, and electron-phonon coupling,

$$H = \sum_{MN} \varepsilon_{MN}\, a_M^\dagger a_N + \sum_Q \hbar\, \omega_Q (b_Q^\dagger b_Q + 1/2) + \sum_{MN,Q} \hbar\, \omega_Q g_{MN}^Q (b_Q^\dagger + b_{-Q}) a_M^\dagger a_N. \quad (1)$$

Here $\varepsilon_{MM}$ and $\varepsilon_{MN}$ are the onsite energies and transfer integrals in a chosen orbital basis. $\hbar\omega_Q$ are the phonon energies with composite index $Q \equiv (\lambda,\boldsymbol{q})$, where $\lambda$ represents the phonon mode index and $\boldsymbol{q}$ the phonon's wave vector, and $g_{MN}^Q$ are the linear (and dimensionless) electron-phonon coupling (EPC) constants.

Performing the phonon trace within the slow/fast separation as described in Refs. [35,45] yields an effective electronic Hamiltonian

$$\bar{H} = \sum_{MN} \bar{\varepsilon}_{MN}\, a_M^\dagger a_N, \quad (2)$$

which is used to compute the transport properties. Vibrational effects enter through renormalized electronic parameters $\bar{\varepsilon}_{MM} = \varepsilon_{MM} + \Delta\varepsilon_{MM} - E_{MM}^{\text{pol}}$ and $\bar{\varepsilon}_{MN} = (\varepsilon_{MN} + \Delta\varepsilon_{MN}) f_{\text{nar},MN}$, where $\Delta\varepsilon_{MN}$ describes dynamic disorder arising from slow modes and follows, to leading order, a Gaussian distribution with variance[35,45]

$$\sigma_{MN}^2 = \sum_Q^{\text{slow}} (\hbar\omega_Q)^2 |g_{MN}^Q|^2 (1 + 2N_Q). \tag{3}$$

The factor $f_{\text{nar}}$ accounts for the polaronic narrowing induced by fast modes. Finally, the charge-carrier mobility $\mu$, in the TPL approach, derived within the Kubo formalism, is obtained as

$$\mu = \frac{e}{2k_B T(\tau_{\text{slow}}^2)} \int_0^\infty e^{-\frac{t}{\tau_{\text{slow}}}} \langle \Delta X^2(t) \rangle dt = \frac{e}{2k_B T} \frac{L_{\text{loc}}^2}{\tau_{\text{slow}}} . \tag{4}$$

$\tau_{\text{slow}}$ is the effective time scale at which the disorder landscape changes due to the slowly moving quasi-static modes. $\langle \Delta X^2(t) \rangle$ is the time-resolved, thermally-averaged quantum spread calculated from the time evolution of a random-phase state[46] based on the Hamiltonian (2). $L_{\text{loc}}^2$ defines the squared localization length, i.e. the average spread of the polaronic wave packet up to the time $\tau_{\text{slow}}$.[35]

Contrary to earlier work,[35] where the electronic basis states were chosen to be orthogonalized versions of frontier molecular orbitals of isolated molecules, the present approach uses Wannier orbitals as the electronic basis states, as detailed further below. Their inherent properties, i.e., orthonormality, spatial localization and adaptation to the frontier states, make them an ideal choice as an orthonormal tight-binding-like basis even for covalently bonded systems.

*2.1.2. Implementation Based on Wannier Orbitals*

All the relevant parameters (such as $\Delta\varepsilon_{MN}$, $f_{\text{nar}}$, $E_{MM}^{\text{pol}}$) required for the TPL simulation of the mobility are calculated from the material-specific quantities in the Hamiltonian (1). The accurate computation of phonon energies and EPC constants has been discussed in the literature, with benchmarks across different theoretical frameworks and density-functional theory (DFT) approximations.[47] Computational details underlying this work are provided in the Methods section below. In the following, we describe how electronic couplings, phonon modes, and EPC constants are obtained within the Wannier-orbital-based framework.

Wannier orbitals can be calculated using the relation[41]

$$|W_{a,l}\rangle = \frac{1}{\sqrt{N_{\boldsymbol{k}}}} \sum_{n\boldsymbol{k}} e^{-i\,\boldsymbol{k}\cdot\boldsymbol{R}_l} |\Psi_{n\boldsymbol{k}}\rangle\, U_{an,\boldsymbol{k}}. \tag{5}$$

Here, $n$ represents the band-index for the Bloch functions $|\Psi_{n\boldsymbol{k}}\rangle$, $\boldsymbol{k}$ is the wave vector in the first Brillouin zone, $a$ indexes the Wannier orbitals in a given unit cell, and $l$ represents the unit cell index for the location of the orbital. $N_{\boldsymbol{k}}$ is the number of wave vectors in the first Brillouin zone, which also corresponds to the number of unit cells in the considered supercell.

Wannier functions are not uniquely defined, since different phase choices of the underlying Bloch states can lead to differently localized real-space orbitals. This freedom can be expressed mathematically through a unitary transformation of the Bloch functions. A common strategy to fix this ambiguity is the construction of maximally localized Wannier functions (MLWFs), in which a spread functional is minimized to obtain the most spatially localized representation.[41,48] An alternative and computationally simpler approach is the projection method, [41] which corresponds to the zeroth-order approximation of the maximal-localization procedure.[49] In the present work, we adopt the projection approach as detailed by Engel et al.[50]. The approach starts with a set of localized trial orbitals $\zeta_a(\boldsymbol{r})$ , which define the centers and approximate shapes of the resulting Wannier functions. In practice, simple atom-centered Gaussian-type functions, or linear combinations thereof, are sufficient. These trial orbitals are projected onto a selected Bloch-state manifold at each wave vector $\boldsymbol{k}$, yielding a projection matrix

$$\boldsymbol{A}_{an,\boldsymbol{k}} \;=\; \boldsymbol{D}_{n\boldsymbol{k}} \, \langle \Psi_{n\boldsymbol{k}} | \zeta_a \rangle. \tag{6}$$

The weight factors $\boldsymbol{D}_{nk}$ can be used to provide a smooth cutoff for the higher-lying Bloch states. $\boldsymbol{A}_{na,\boldsymbol{k}}$ is then expressed as a matrix product using singular value decomposition,

$$\boldsymbol{A}_{na,\boldsymbol{k}} = \sum_m^{N_b} \sum_b^{N_W} X_{nm,\boldsymbol{k}} \, \Lambda_{mb,\boldsymbol{k}} Y^{\dagger}_{ba,\boldsymbol{k}}. \tag{7}$$

Finally, replacing the matrix of singular values $\Lambda_{mb,\boldsymbol{k}} \to \delta_{mb}$, we obtain the desired unitary transformation

$$U_{na,\boldsymbol{k}} = \sum_m^{N_b} \sum_b^{N_W} X_{nm,\boldsymbol{k}} \, \delta_{mb} Y^{\dagger}_{ba,\boldsymbol{k}}. \tag{8}$$

to calculate Wannier orbitals from the DFT-Bloch states. The resulting projected Wannier functions are orthonormal, moderately localized, and accurately reproduce the frontier electronic states (*vide infra*). With the Wannier-orbital basis $|W_M\rangle$ fixed, and $M$ a shorthand notation for the Wannier-orbital index $M \equiv (a, l)$, the electronic couplings (onsite energies and transfer integrals) between the orbitals are calculated as

$$\varepsilon_{MN} \;=\; \langle W_M | H | W_N \rangle, \tag{9}$$

in which $H$ represents the Kohn-Sham (KS) Hamiltonian from the DFT simulation of the material.

The vibrational normal modes for the material are obtained by numerical diagonalization of the mass-weighted Hessian with the DFT simulations carried out in CP2K.[47] Following previous work[51,52], we define the EPC constant in the Wannier representation as

$$g_{MN}^{Q} = \frac{1}{\sqrt{2\hbar\omega_Q^3}} \left\langle W_M \middle| \frac{\partial H}{\partial X_Q} \middle| W_N \right\rangle, \tag{10}$$

where the derivative of the Kohn-Sham Hamiltonian $H$ with respect to the normal mode coordinates $X_Q$ is performed numerically yielding [53]

$$\frac{\partial H(\mathbf{R})}{\partial X_Q}\Big|_{\mathbf{R}^0} \approx \frac{H(\mathbf{R}^0+\delta\mathbf{e}_Q)-H(\mathbf{R}^0-\delta\mathbf{e}_Q)}{2\delta\sqrt{m_Q}}. \tag{11}$$

Here $\mathbf{e}_Q$ represents the normalised polarisation vector and $m_Q$ is the mass associated with the normal mode $Q$ and the coordinates $X_Q$. We note that both local ($M = N$) and non-local ($M \neq N$) EPC constants and energetic fluctuations are evaluated consistently on the same microscopic footing, and the influence of the crystalline environment is inherently included in the electronic structure and its vibrational modulation.

We performed all the DFT calculations within the generalized gradient approximation (GGA) using the PBE functional[54] and the CP2K[55] code. For complete details of these simulations, we refer to the Methods section, while we focus here on the relevant aspects. The DFT simulations for the calculation of EPC were performed using a supercell approach. This strategy provides a convenient and consistent way to access both electronic Bloch states and phonon modes on equivalent grids in reciprocal space. A supercell calculation at the Γ- point implicitly corresponds to sampling a Γ -centered k-grid in the primitive Brillouin zone for the electronic states, while simultaneously enabling access to the corresponding q-grid for lattice vibrations if needed. Owing to Brillouin-zone folding, the electronic eigenstates obtained from the supercell Γ-point calculation span the full set of states associated with the underlying primitive-cell k-grid. However, because states related by time-reversal symmetry are degenerate, the eigenvectors provided by the DFT supercell calculation are, in general, arbitrary linear combinations within these degenerate subspaces and therefore do not necessarily correspond to well-defined Bloch states of the primitive lattice. To recover Bloch states with well-defined crystal momentum, we apply a unitary transformation within each degenerate subspace that diagonalizes the translation operators of the primitive lattice. Since Bloch states are simultaneous eigenstates of the Hamiltonian and the translation operators, this procedure uniquely restores the Bloch character of the electronic states. The resulting Bloch states $|\Psi_{n\mathbf{k}}\rangle$ are then used as input for the Wannierization procedure in Eq. (5).

## 2.2. Material and Structure

We apply the above implementation for the simulation of electron and hole transport in the widely studied block copolymer P(NDI2OD-T2) *(poly(N,N'-bis-2-octyldodecylnaphthalene-1,4,5,8-bis-dicarboximide-2,6-diyl-alt-5,5-2,2-bi-thiophene)*[42]. This highly soluble and air-stable polymer[56] has been used for several applications, such as organic field-effect transistors (OFETs)[57,58] and also for organic photovoltaics (OPVs)[59,60]. This material can crystallize into three polymorphs (Form I-III), which differ in π–stacking motif and in the relative orientation of the polymer chains to the substrate[57,58,61,62]. While these polymorphs are relevant for thin-film device performance, the present study does not aim to resolve polymorph-specific transport properties. Instead, we focus on generic structural motifs representative of the π–π stacking and backbone connectivity in P(NDI2OD-T2). This provides a controlled setting to identify intrinsic microscopic transport mechanisms and, in particular, to disentangle transport along the conjugated backbone from interchain transport across van der Waals gaps.

For this purpose, we consider the following simplified model of P(NDI2OD-T2): We reduce the alkyl side-chains to a single hydrogen atom in favor of reducing computational expense. In conjugated polymers, alkyl side chains primarily affect solubility and processing and do not directly participate in the backbone's π-states, but affect electronics indirectly through conformation and packing[63–65]. Because the H-substituted model preserves the P(NDI2OD-T2) repeat unit and bonding pattern, we expect the backbone electronic structure to remain largely governed by the aromatic core. Modest side-chain–induced conformational differences cannot be excluded but are not expected to change our conclusions at the level considered here. We therefore refer to this model as P(NDIH-T2). Consistent with this simplification, the electronic coupling between polymer stacks *across the side-chain direction* are not considered.

To obtain a ground-state crystal structure, we perform geometry optimization with a unit cell containing a stacked dimer of P(NDIH-T2) (two chains per unit cell) with periodic boundary conditions in all directions. We constrain the optimization to an orthorhombic unit cell with a sufficiently large vacuum (~ 20 Å) in the direction where alkyl side-chains would separate polymers, hence suppressing any interaction in this direction.

Figure 1 (a) and (b) show the resulting ground-state crystal structure. The unit cell axis $\boldsymbol{a}$ is close to the long axis of the NDI moiety (also pertaining to direction of the here-reduced alkyl side-chains), $\boldsymbol{b}$ is along the $\pi -$stacking direction and $\boldsymbol{c}$ is along the polymer backbone. The lattice parameters are $a = 25.0$ Å, $b = 7.9$ Å and $c = 14.2$ Å. The dihedral angle between the NDI moiety and bithiophene moiety is about 47°. We note that, as seen in Figure 1 (b), the polymer chains are slightly shifted along $\boldsymbol{c}$-direction. The corresponding CIF file for this simulated structure is provided in the Supporting Information (SI).

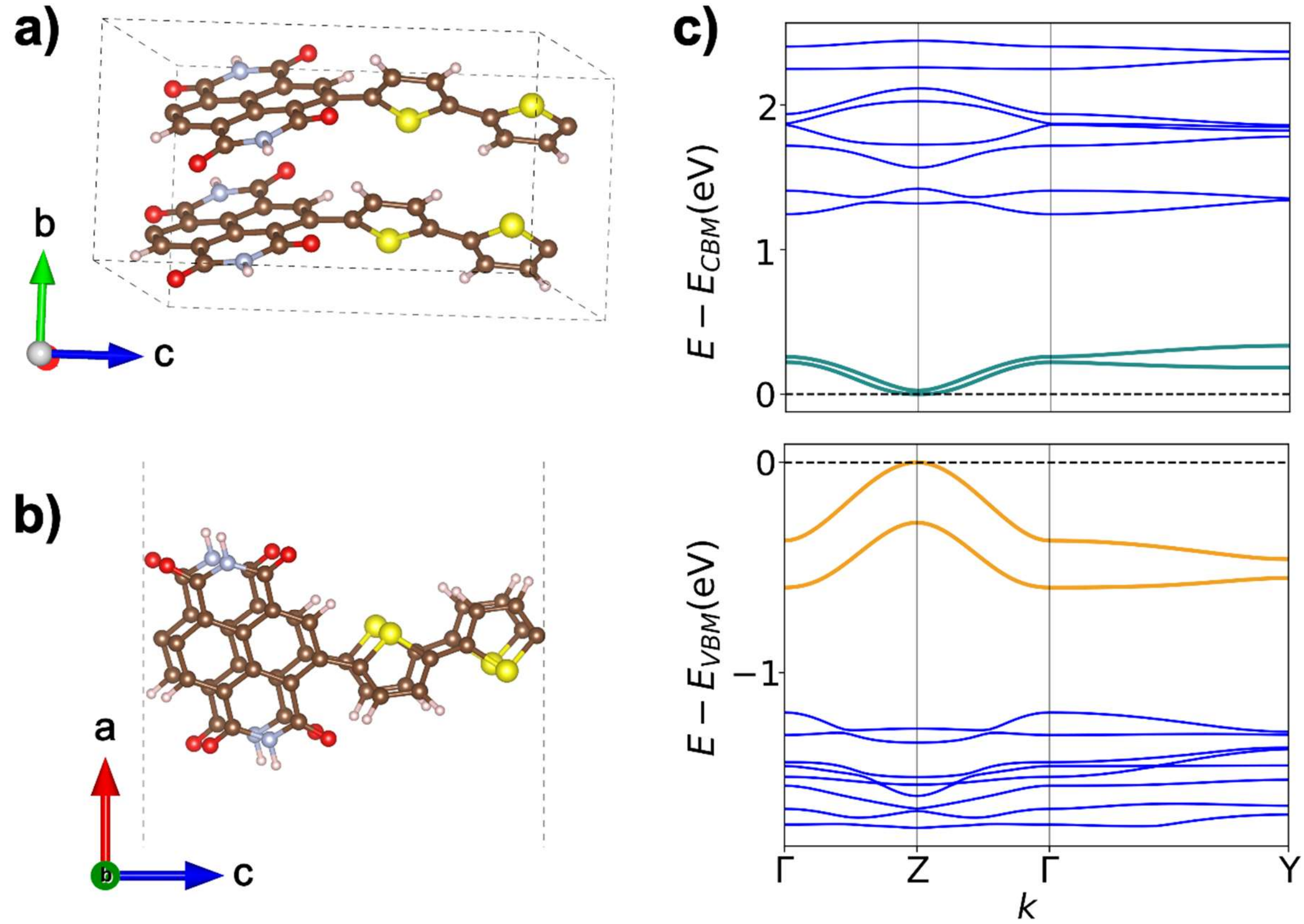

***Figure 1: Calculated ground-state crystal structure and band structure of P(NDIH-T2)*** *(a) as viewed along* $\boldsymbol{a}$ *(close to NDI long-axis direction) (b) as viewed along* $\boldsymbol{b}$ *(*$\pi$ *−stacking direction). (c) Band structure of P(NDIH-T2) simulated at DFT/PBE level of theory. The special points in the Brillouin zone are labelled as* $\Gamma = (0,0,0)$*,* $Z = (0,0,0.5)$ *and* $Y = (0,0.5,0)$ *in terms of the reciprocal lattice vectors. Energy zero is chosen at the valence band maximum (VBM) following standard convention.*

Relating our calculated structure with experimental findings, we obtain a $\pi$ −stacking distance of 3.95 Å (approximated as half of the lattice parameter $b$), while reported experimental values lie in the range $3.81 - 3.93$ Å[57,58,66]. This comparison is particularly relevant because the $\pi$ stacking distance strongly affects the transfer integral in the stacking direction. Similarly, the unit cell lattice parameter $c$ is 14.2 Å for the calculated structure, compared to reported experimental values of $13.9 - 14.7$ Å[58,61,66] although the corresponding unit cells are not orthorhombic. Moreover, Karunasena et al.[61] reported a related polymorph, denoted Form I-β, in which the polymer chains also exhibit a shift along the $\boldsymbol{c}$-direction. Given the structural variability across experimentally reported polymorphs, we do not aim at a one-to-one reproduction of a specific experimental structure. Rather, these comparisons indicate that the present model captures the relevant local packing motifs and structural length scales needed for the transport analysis.

### 2.3 Band Structure and Wannier Orbitals

To investigate the electronic properties, we calculate the band structure along the principal transport directions (Figure 1 (c)). Here, the $\Gamma - Z$ path represents the polymer backbone direction while the $\Gamma - Y$ path represents the $\pi -$stacking direction. We find that both the valence and conduction bands exhibit the highest dispersion along the polymer backbone, with bandwidths of 372 meV and 222 meV, respectively. In contrast, the dispersion in the $\pi$-stacking direction with 88 meV and 39 meV, respectively, is significantly smaller. This type of anisotropy has been observed previously for several other $\pi -$conjugated polymers (such as P3HT, PBTTT and PQT-12)[64,67–69], and could be expected from chemical intuition due to stronger delocalization of the electronic states along the polymer backbone arising from $\pi$ conjugation[70].

Interestingly, in a recent study on P(NDI2OD-T2)[61], it was reported that the anisotropy follows another trend for Form I$-\beta$(anti) (i.e. the structure with a shift of the polymer stacks along $\boldsymbol{c}$), for which the conduction band width along the polymer chain was reported to be 309 meV, whereas that across the $\pi -$stacks was 535 meV. While the bandwidth along the polymer backbone is of similar order to our calculations, that in the $\pi -$stacking direction is much larger. This is attributed to variations in overlap of the orbitals of neighboring chains. That is, the $\pi$ stacking fashion considered in their work for Forms I$-\alpha$ and I$-\beta$ are inclined structures and contain only one polymer chain per unit cell. This simpler stacking model favors larger overlap and electronic coupling, while our structure corresponds to serrated stacking with two chains that are laterally displaced and with roughly doubled unit cell in $\boldsymbol{c}$ direction. We also note that the corresponding transfer integrals in the stacking direction change sign between Forms I$-\alpha$ and I$-\beta$ in Ref.[61] This sign-change means that any structure intermediate between these forms could therefore have significantly smaller transfer integrals[71] in full consistency with our result.

Since P(NDI2OD-T2) is commonly used as an n-type semiconductor, it is also interesting to note that the valence band dispersion is significantly larger than that of the conduction band in both directions. Hence, it will be interesting to compare the transport of both charge carriers in this material further below. In the following, we therefore analyze transport for the two energetically lowest bands of each carrier type, specifically represented in orange (holes) and blue (electrons) in Figure 1(c). Henceforth, we refer to the underlying Wannier orbitals that are derived from the valence bands as hole Wannier orbitals and those from the conduction bands as electron Wannier orbitals.

Bloch states from the DFT simulations are obtained within $1 \times 3 \times 3$ supercell of the material, and Wannierization was performed following the methodology detailed in Section 2.2. For the projection step, localized trial orbitals were chosen as a linear combination of atom-centered Gaussian-type orbitals (GTOs) centered on chemically relevant regions of the molecule (thiophene units for holes and NDI cores for electrons). We emphasize, however, that these trial orbitals merely define the initial projection and do not represent the final Wannier functions. The resulting orbitals are defined by the underlying Bloch states and emerge from the projection procedure as orthonormal, physically meaningful localized states.

In particular, the obtained Wannier orbitals for the valence bands extend over two adjacent thiophene units (cf. Fig. 2 (a)), indicating that the relevant electronic states are not confined to individual rings but delocalized over larger molecular segments. This illustrates that the proposed Wannier orbital construction naturally identifies the appropriate localization length scale associated with a selected band manifold, without requiring an *a priori* choice of molecular fragments. Consequently, we obtain two valence Wannier orbitals per unit cell, each localized on a bithiophene bridge within the stack.

We next construct two separate tight-binding models for electrons and holes, based on their respective Wannier orbitals. For better visualizing the model, in Figure 2 (a), we summarize schematically the largest transfer integrals in the hole tight-binding model. Figure 2 (b) and (c) display one of the hole and electron Wannier orbitals, respectively, in the 1x3x3 supercell.

Finally, to ultimately test the Wannier-orbital description we reconstruct the Bloch Hamiltonian, i.e., essentially performing a Wannier interpolation of the electronic structure in $k$-space, and plot the resulting band structure in comparison to the original DFT-band structure. As shown in Figure 2 (d) and (e), a direct comparison in the zoomed energy window confirms that the Wannier-interpolated bands closely track the DFT dispersion. Note that from this compact Wannier description, we directly obtain the associated transfer integrals, which provide direction-resolved measures of electronic coupling and thus serve as indicators of charge-transport anisotropy. The maximum transfer integrals along the crystallographic $\boldsymbol{b}$- and $\boldsymbol{c}$-directions are summarized in Table 1 (first column). Consistent with the DFT band structure, a pronounced difference between charge carriers and anisotropy in electronic couplings is reflected in the corresponding values of $\varepsilon_{MN}^{\max}$ (maximum transfer integrals in the respective direction).

| | Direction | $\varepsilon_{MN}^{\max.}$ (meV) | $\sigma_{MN}$ (meV) | $\sigma_{MM}$ (meV) | $\tau_{\text{slow}}$ (fs) | $L_{\text{loc}}$ (Å) | $\mu$ (cm$^2$/Vs) |
|---|---|---|---|---|---|---|---|
| hole | $\boldsymbol{c}$ | 88 | 13 | 70 | 58 | 36.0 | 42.7 |
| | ***b*** | 44 | 34 | | | 6.8 | 1.5 |
| electron | ***c*** | 54 | 5 | 43 | 61 | 41.5 | 54.3 |
| | ***b*** | 47 | 17 | | | 7.3 | 1.7 |

**Table 1:** Simulated material parameters and calculated localization lengths and mobilities for electrons and holes in different directions.

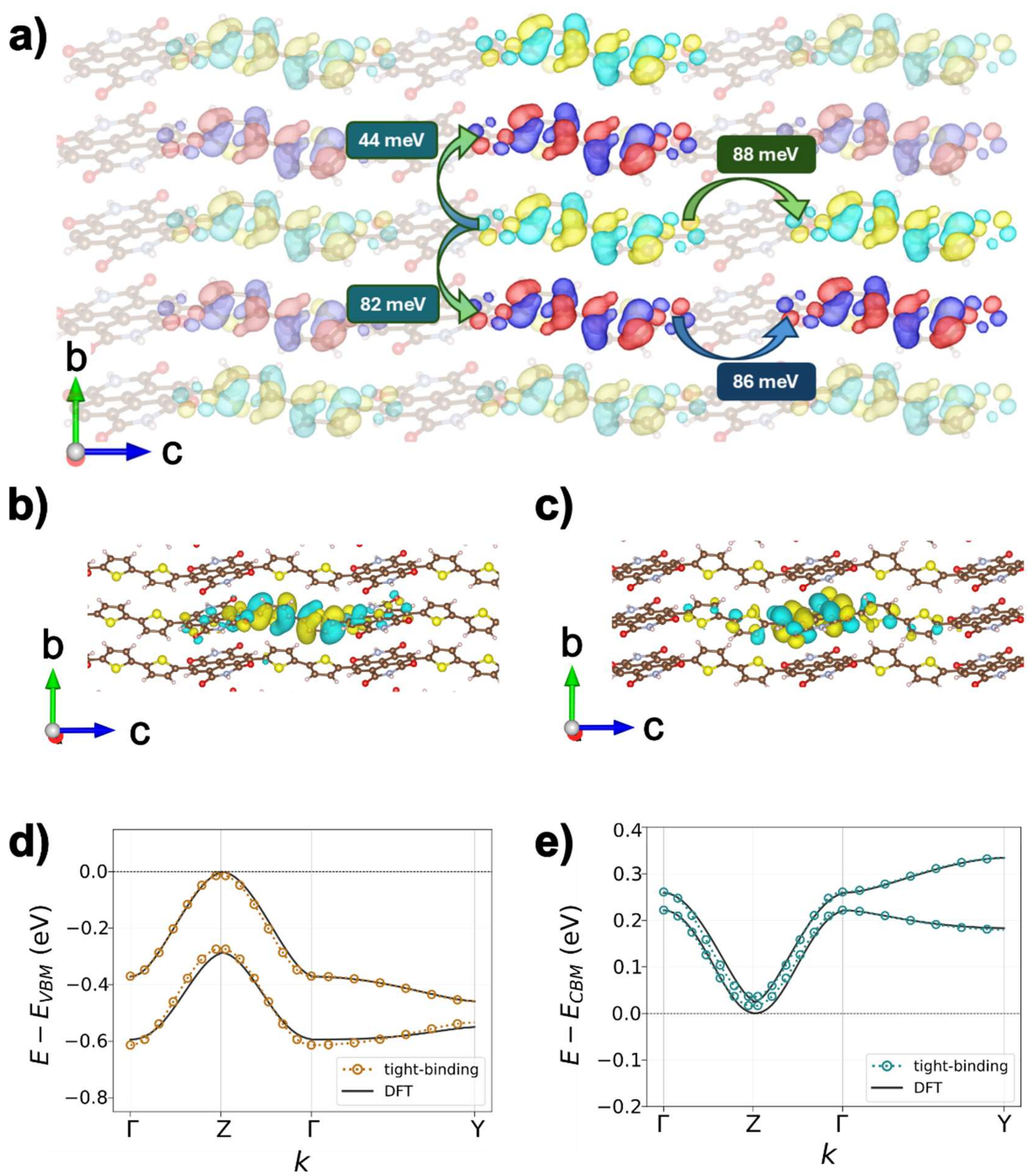

***Figure 2: Wannierisation of the low-energy band structure.*** *(a) Visualisation of the Wannier tight-binding model exemplarily for valence states. The Isosurface value was set to 0.03* $(bohr)^{-3/2}$*. (b,c) Plots of one of the Wannier functions for the valence (b) and conduction (c) manifold each (isosurface of 0.02* $(bohr)^{-3/2}$*). Comparison of Wannier interpolated bands (red dots) with original DFT band structure (blue lines) for valence (d) and conduction bands (e) respectively. VBM = Valence Band Maximum and CBM = Conduction Band Minimum.*

### 2.4. Vibrations and electron-phonon coupling

To reduce the computational effort – and given that the unit cell already contains two polymer chains in the stacking direction as well as three repeat units along the chains – we restrict the vibrational analysis to $\Gamma$ point phonons ($\boldsymbol{q}$ =(0,0,0)) of the primitive phononic BZ, resulting in 225 phonon modes. Because of the two T2–NDI chains stacked along $\boldsymbol{b}$, the $\Gamma$-point treatment effectively captures both in-phase and anti-phase distortions, corresponding to $\boldsymbol{q} = (0,0,0)$ and $q = (0, \frac{\pi}{b}, 0)$ in a single-chain cell. Thus, both intra-chain vibrations and inter-chain relative motions are captured, the latter arising from the anti-phase combinations of the stacked chains. Along the chain direction $\boldsymbol{c}$, the unit cell is already extended and contains several internal dihedral degrees of freedom within a single cell. Consequently, $\Gamma$-point phonons already incorporate substantial backbone flexibility, reducing the need for additional $\boldsymbol{q}$-point sampling in this direction. In principle, the supercell DFT framework would also permit the inclusion of phonon modes associated with finer $\boldsymbol{q}$-point grids of the chosen supercell. However, this would increase the computational cost by roughly an order of magnitude, rendering such calculations presently impractical for the system under study. For the selected set of 225 modes, we compute the EPC constants $g_{MN}^{Q}$ using the method described in Sec. 2.2. To study the effect of vibrations on the electronic structure, we calculate the quasi-static disorder $\sigma_{MN}^2$ and the polaronic narrowing $f_{\mathrm{nar},MN}$, both of which account for the impact of the vibrations in the charge transport model.

**Figure 3** compares $\sigma_{MN}^2$ and $f_{\mathrm{nar},MN}$ for electron and holes across different types of electronic couplings, including onsite, intra-unit-cell, inter-unit-cell (along stacking direction and along the backbone direction) as depicted in the corresponding sketches on the left. To ensure a fair comparison between both carrier types, we select the pair of Wannier orbitals with the highest transfer integral to represent each of these categories. Specifically, we plot the cumulative variance of the quasi-static disorder by incrementally adding contributions from phonon modes with increasing energies, up to the separation energy (represented by the dashed lines) that distinguishes the low- and high-frequency treatment in our model according to the criterion introduced above ($87.7\ meV = 707.9\ cm^{-1}$ for holes and $54.1\ meV = 436.3\ cm^{-1}$ for electrons). Likewise, we depict the cumulative polaronic narrowing, obtained by consistently adding contributions from the highest-energy phonon modes and with decreasing energy.

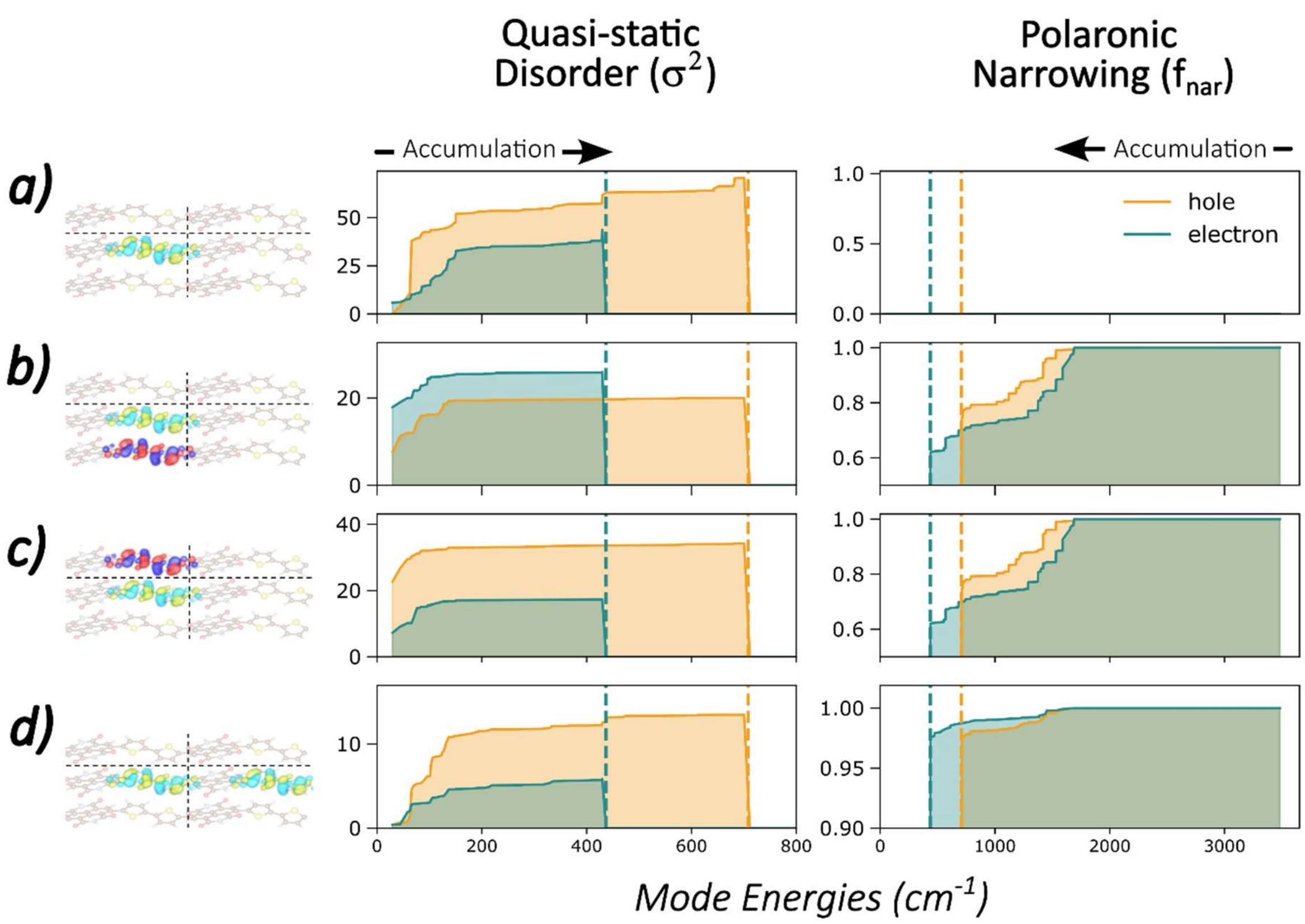

***Figure 3: Cumulative electron-phonon-coupling effects.*** *Cumulative quasi-static disorder from low to high energies (accumulation direction indicated by arrow) and cumulated polaronic narrowing from high to low energies (accumulation direction indicated by arrow); Both quantities are compared for pairs of Wannier orbitals with different index combinations, visualized in the first column: (a) for the same Wannier orbital (corresponding to a local coupling constant); (b) between a pair of Wannier orbitals in the same unit cell; (c) between an orbital pair in neighbouring unit cells in the* ***b****-direction; and (d) between a pair of Wannier orbitals in neighbouring unit cells in the* ***c****-direction. Legend applies to all panels. Different coloring scheme for orbitals indicates different Wannier orbitals.*

Several notable features of P(NDIH-T2) are observed from Figure 3. Firstly, the quasi-static disorder is strongly determined by the lowest of the low-frequency modes (mostly below 150 $\mathrm{cm}^{-1}$), while modes in the range 200-400 $\mathrm{cm}^{-1}$ are hardly influential. Secondly, holes typically experience higher quasi-static disorder compared to electrons in nearly all instances. Specifically, holes exhibit quasi-static disorder that is 1.6, 2 and 2.6 times greater for the onsite and non-local coupling along the stacking and backbone directions, respectively (Fig. 3 a,c,d), while, non-local couplings between electron orbitals within the unit cell experience a quasi-static disorder that is 1.3 times greater than that of holes (Fig. 3b). In consistency to the low-frequency range we can attribute a larger $\sigma_{MN}^2$ for holes to the higher flexibility of the bithiophene unit. This quantitatively different effect of vibrations on electrons vs. holes is summarized in Table 1.

Additionally, it can also be seen that in the stacking direction, the electrons experience a slightly stronger polaronic narrowing of the transfer integrals (right panels in Fig. 3). One of the most significant contributions to the narrowing arises from the phonon modes around 1589 $cm^{-1}$. These modes are identified as the C-C stretch bonds within the NDI moiety. Since the electronic Wannier orbitals are predominantly located on the NDI moiety (as seen in Figure 2), while the hole Wannier orbitals are predominantly located on the bithiophene moiety, it is understood why the electrons experience stronger polaronic narrowing due to this phonon mode.

### 2.5. Charge Transport

After determining all relevant material parameters, we perform charge-transport simulations for both electrons and holes. We first analyze transport in P(NDIH-T2) within the TPL framework for the two crystallographically relevant directions, namely along the polymer backbone and along the $\pi$ -stacking directions. This full crystalline model is referred to in the following as the bulk model and serves to quantify the transport anisotropy.

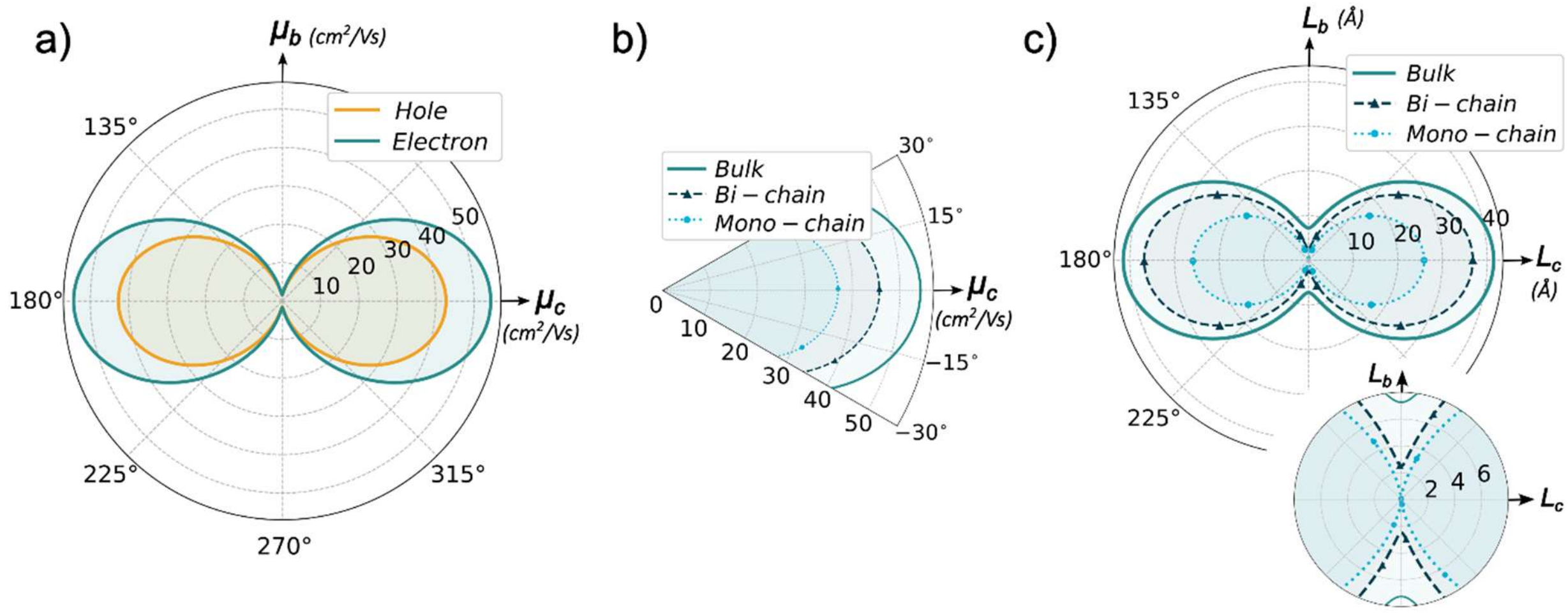

***Figure 4: Charge carrier mobilities and localization lengths of P(NDIH-T2).*** *(a) Direction dependence of carrier mobility of electrons and holes in* $\boldsymbol{b}-\boldsymbol{c}$ *plane of the full (bulk) model;* $\boldsymbol{c}$*-direction is the polymer backbone direction, cf. Fig. 2; (b) electron mobilities for the bulk model, bi-chain and mono-chain model systems close to the* $\boldsymbol{c}$*-direction; (c) electron localization lengths* $L_{loc}$ *for the bulk, bi-chain and mono-chain cases. Inset shows a zoomed in area near the origin, highlighting the anisotropy for small* $L_{loc}$ *values.*

**Figure 4** (a) displays the calculated mobilities extracted from the simulated spreading dynamics; the corresponding numerical values are summarized in Table 1. We first discuss the anisotropy, which is a central feature of charge transport in organic solids. While the mobilities reach decently high values of about 1.5 $cm^2/(Vs)$ or above along the π-stacking direction (***b***-direction), they are even much higher in polymer backbone direction (***c***-direction), indicating significantly more efficient transport along the chain. In this direction we observe values of 42.7 and 54.3 $cm^2/(Vs)$ for holes and electrons, respectively, which yields anisotropy factors of about 30 for both carrier types.

The strong anisotropy has several microscopic origins. Firstly, the direction dependence in the transfer integrals as discussed above would directly impact on the mobility anisotropy, however as analyzed in Fig. 2, this direction dependence remains comparatively moderate. Second, for both charge carrier types, vibrational disorder is significantly stronger along the π-stacking direction than along the polymer backbone, as apparent from comparing Fig. 3c, d. This trend is also reflected in the localization lengths $L_{\mathrm{loc}}$ in Fig. 4c, which quantify the average spatial extent of the transiently localized polarons within the TPL framework. $L_{\mathrm{loc}}$ values are significantly shorter in the π-stacking than in the polymer backbone direction, indicating stronger confinement of the charge carriers. Finally, the larger lattice constant along the polymer backbone has also a substantial impact. Together, these contributions eventually lead to mobilities that are approximately 30 times higher along the polymer backbone similarly for both carrier types.

Comparing electron and hole transport, we find that the electron mobilities are, on average, about 18% higher than those of holes. Although the transfer integrals alone would have suggested a higher hole than electron mobility, the substantially stronger electron–phonon coupling for holes observed and analyzed above overcompensates the transfer integral difference. This difference is also directly reflected in the localization lengths. In particular, electrons exhibit systematically larger localization lengths (up to 41.5 Å, corresponding to three repeat units, cf. Fig. 4(c)) than holes.

The markedly higher mobilities along the backbone discussed above raises the question of whether transport in this direction can be understood within a single-chain picture, or whether inter-chain processes provide a significant contribution for chain-parallel transport. To address this, we perform additional charge-transport simulations on two modified models. In the first one, referred to as the *"bi-chain model"*, we retain two adjacent polymer chains from the full system. All intra-chain interactions are preserved, while inter-chain transfer integrals and electron–phonon couplings are restricted to this pair of chains, i.e., interactions with additional neighboring chains are suppressed. In contrast, the second model, termed the "*mono-chain* *model"*, consists of a single isolated polymer chain with all inter-chain interactions removed. The results of both models are summarized and compared to the bulk solid in Fig. 4(b) and (c).

Within the mono-chain model, the mobilities of electrons and holes decrease to 37.0 $\mathrm{cm^2/(Vs)}$ and 25.4 $\mathrm{cm^2/(Vs)}$, respectively, which correspond to only about 64% percent of the bulk values. This result demonstrates that charge transport along the polymer backbone, although predominantly intra-chain in character, is not fully captured by an isolated-chain picture. Inter-chain electronic coupling provides an important additional contribution, because it allows the charge carrier to transiently delocalize onto neighboring chains. This temporary access to adjacent chains enhances the overall transport efficiency along the backbone and increases the effective intra-chain mobility by a factor of about 1.6 on average. This effect emerges already in the case of two stacked polymer chains as probed in the *bi-chain model,* which yields mobilities intermediate between those of the bulk system and the isolated chain.

In full consistency, the localization lengths along the backbone reduces in the bi-chain and mono-chain models (cf. Fig. 4c). For electrons in the single chain, $L_{loc}$ reduces from three to about two repeat units (25.9 Å). An analogous trend is found for holes, reflecting the same supporting role of transient access to adjacent chains.

## 3. Discussion and Conclusion

Experimental studies on bar-coated P(NDI2OD-T2) thin films, in which the polymer chains are highly aligned along the printing direction, report field-effect mobilities of up to 6.4 $\mathrm{cm^2/(Vs)}$ in the saturation regime and 1.2 $\mathrm{cm^2/(Vs)}$ in the linear regime, along with a pronounced transport anisotropy of approximately 35 between directions parallel and perpendicular to the printing axis in the saturated regime and 15 in the linear regime.[72] This observation is consistent with the anisotropy values obtained in our simulations. In particular, our results confirm that charge transport is predominantly governed by motion along the polymer backbone, while inter-chain electronic coupling along the π-stacking direction provides an essential supporting mechanism. By enabling transient transverse polaron delocalization, inter-chain pathways enhance effective intra-chain transport and contribute to the observed anisotropy.

A pronounced asymmetry between electron and hole mobility has also been reported experimentally, with hole mobilities approximately an order of magnitude lower than those of electrons. [73] This trend is in qualitative agreement with our theoretical results and shows that a clear carrier asymmetry already emerges from the intrinsic crystalline transport regime captured by the present model. In particular, the calculations indicate that electrons and holes are affected differently by the interplay of electronic coupling and vibrational disorder, leading to stronger localization and lower mobility for holes. At the same time, in experimental thin-film samples the observed asymmetry may be further influenced by carrier-specific defects or trap states, which are beyond the scope of the present crystalline model.

Beyond the specific case of P(NDI2OD-T2), the present work extends the TPL framework to crystalline π-conjugated polymers by formulating all relevant Hamiltonian parameters in a Wannier-orbital representation. This avoids the need for predefined molecular fragments and provides a consistent basis for describing charge transport in extended conjugated systems. In this way, intra- and inter-chain transport processes can be treated on the same footing, allowing their interplay to be analyzed systematically. The approach is not restricted to the present material and should be applicable to a broader class of crystalline organic semiconductors.

## 4. Experimental Section/Methods

*Density functional theory simulations:* For all material parameters and the Hamiltonian parametrization, density functional theory calculations were performed within the generalized gradient approximation (GGA). We used the PBE functional[54] with the GTH pseudopotentials[74–76] and TZV2P basis set (from the BASIS-MOLOPT collection)[77] along with periodic boundary conditions as implemented in CP2K[55]. Dispersion interactions were taken into account using Grimme's DFT-D3 method[78,79]. Geometry optimizations were performed in the primitive unit cell until, where the maximum geometrical change (atomic positions and lattice parameters) between consecutive steps was below $2 \times 10^{-3}$ bohr ($\approx 10^{-3}$ Å) and the maximum forces on individual atoms were smaller than $1 \times 10^{-5}$ hartree/bohr ($\approx 5 \times 10^{-4}$ eV/Å). Band structure calculations were performed in the primitive unit cell using a Monkhorst-Pack[80] k-mesh with the grid size of $1 \times 5 \times 3$.

**Acknowledgements**

We would like to thank the Deutsche Forschungsgemeinschaft for financial support [projects 511287670, 541495916, TRR 386 (TP B05, project number 514664767), and the Cluster of Excellence e-conversion (Grant No. EXC 2089/2-390776260)]. We gratefully acknowledge grants for computer time from the Leibniz Supercomputing Centre in Garching and for computing time made available on the high-performance computer Barnard at the NHR Center TUD-ZIH. This center is jointly supported by the Federal Ministry of Education and Research and the state governments participating in the National High-Performance Computing (NHR) joint funding program (http://www.nhr-verein.de/en/our-partners).

**Data Availability Statement**

All data supporting the findings of this study are contained in the manuscript and are available from the corresponding author upon reasonable request.

**Code Availability Statement**

Custom python scripts used in the study are available from the corresponding author upon reasonable request.

**Author Contributions:**

N.T. : Investigation (Performing DFT calculations and transport simulations), Coding, Data analysis and visualisation, writing - original draft. M.F.D : Coding, Data analysis, writing. M.P. : Coding, Data analysis,writing. F.O. : Conceptualisation, Supervision, Funding acquisition, writing. All authors contributed to reviewing and editing and approved the manuscript.